**Enhancing strength of MICP-treated sandy soils: from micro to macro scale**

Yuze Wang[*1]; Charalampos Konstantinou[2]; Kenichi Soga[3]; Jason T. DeJong[4]; Giovanna Biscontin[2]; Alexandre J. Kabla[2]

[1]Department of Ocean Science and Engineering, Southern University of Science and Technology,518055, People's republic of China
[2]Department of Engineering, University of Cambridge, Cambridge, CB2 1PZ, United Kingdom
[3]Department of Civil and Environmental Engineering, University of California, Berkeley, CA 94720, United States
[4]Department of Civil and Environmental Engineering, University of California, Davis, CA 95616, United States
*Correspondence: wangyz@sustech.edu.cn; ORCID: 0000-0003-3085-5299

**Abstract** Microbial-Induced Calcium carbonate ($CaCO_3$) Precipitation (MICP) has been extensively studied for soil improvement in geotechnical engineering. The properties of calcium carbonate crystals such as size and quantity affect the strength of MICP-treated soil. This study demonstrates how the data from micro-scale microfluidic experiments that examine the effects of injection intervals and concentration of cementation solution on the properties of calcium carbonate crystals can be used to optimise the MICP treatment of macro-scale sand soil column experiments for effective strength enhancement. The micro-scale experiments reveal that, due to Ostwald ripening, longer injection intervals allow smaller crystals to dissolve and reprecipitate into larger crystals regardless of the concentration of cementation solution. By applying this finding in the macro-scale experiments, a treatment duration of 6 days, where injection intervals were 12 h, 24 h, and 48 h for cementation solution concentration of 0.25 M, 0.5 M and 1.0 M, respectively, was long enough to precipitate crystals large enough for effective strength enhancement. This was indicated by the fact that significantly higher soil strength and larger crystals were produced when treatment duration increased from 3 days to 6 days, but not when it increased from 6 days to 12 days.

**Keywords:** Soil stabilisation, ground improvement, upscaling, microscopy, optimisation

# Introduction

Microbial-Induced Calcium carbonate ($CaCO_3$) Precipitation (MICP) has been extensively studied for its potential use for soil improvement in geotechnical engineering (van Paassen 2009; Cheng et al., 2012 and 2017; DeJong et al, 2006, 2010 and 2013; Jiang et al., 2017). Ureolysis-driven MICP is among the mostly studied MICP processes due to its ease of control and high chemical transform efficiency (Dhami et al., 2013). Ureolytic bacterial suspensions are injected into soil and attach to particle surfaces, followed by the injection of cementation solution composed of urea and $CaCl_2$. The bacteria hydrolyse urea, producing $CO_3^{2-}$, which reacts with $Ca^{2+}$ to form $CaCO_3$ (equations 1 and 2). The precipitated $CaCO_3$ bonds soil particles and increases the strength of soil matrices.

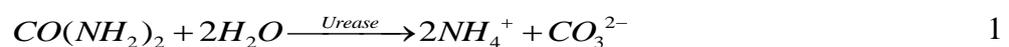

$$CO(NH_2)_2 + 2H_2O \xrightarrow{Urease} 2NH_4^+ + CO_3^{2-} \qquad 1$$

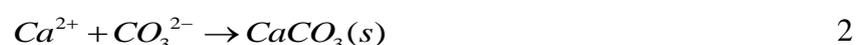

$$Ca^{2+} + CO_3^{2-} \rightarrow CaCO_3(s) \qquad 2$$

Large variations in strength of MICP-treated soil occur even at similar $CaCO_3$ contents, due to of the heterogeneity of $CaCO_3$ crystals (Al Qabany et al., 2012; Cheng et al., 2017). Therefore, it is essential not only to study the factors that affect $CaCO_3$ content, but also the properties of $CaCO_3$ precipitates. The amount of $CaCO_3$ produced in the soil is affected by both the amount of $CaCl_2$ and urea injected into the soils and the chemical efficiency. Chemical efficiency is defined as the percentage of measured mass of $CaCO_3$ relative to the calculated mass of $CaCO_3$ if all the $CaCl_2$ injected into the soil pores transformed into $CaCO_3$ (Al Qabany et al., 2012).

The chemical efficiency depends on the precipitation rate of $CaCO_3$, the concentration of $CaCl_2$ and urea, as well as on the retention period, which is the time interval between two successive injections of cementation solution, i.e. the time given for the urea and $CaCl_2$ to precipitate. At certain cementation solution concentrations, retention period needs to be long enough, or the precipitiation rate needs to be high enough to achieve high chemical efficiency. Al Qabany et al. (2012) showed that, for a bacterial optical density ($OD_{600}$) between 0.8 and 1.2, the chemical efficiency remained high (80%-100%) when the cementation solution injection rate was below 0.042 mole/l per hour, whereas the chemical efficiency decreased when the injection rate exceeded 0.042 mole/l per hour. The input rate was varied by changing the concentration of cementation solution and the retention period.

Whilst the chemical efficiency remained similar, the chemical properties of $CaCO_3$ crystals produced were different when the concentration of cementation solution was altered. The crystals were smaller and coated the sand particles when the concentration was 0.25 M, whereas the crystals were larger and filled the pores when the concentration was 1.0 M (Al Qabany et al., 2012). This implies that changing the treatment process may affect the strength of the MICP-treated sand. In a subsequent work of Al Qabany and Soga (2013), the strength of MICP-treated sand decreased with the increase of cementation solution from 0.25 M to 1.0 M, which was largely due to the inhomogeneity of the $CaCO_3$ distribution in the soil matrix when the concentration of cementation solution was higher.

The reason why crystal properties vary after MICP treatment was largely unknown until recent work conducted by the present authors (Wang et al., 2019b), in which a transparent microfluidic chip designed based on a 2-D cross-sectional image of real soil (Figure 1, Wang et al., 2019a) was used to observe the time-dependent MICP process. Unstable $CaCO_3$ crystals dissolved at the expense of the growth of more stable crystals after the first and second injections of 0.25 M cementation solution. Furthermore, crystals were smaller but larger in number when cementation solution was injected only once per day over a period of 12 days, whereas crystals were larger but smaller in number when the cementation solution was injected two to four times per day. The crystal size may vary depending on whether there is enough time for the dissolution and re-precipitation process to occur, and even when the concentration of cementation solution is the same, the interval between cementation solution injections may affect the size of $CaCO_3$ crystals, thereby affecting the strength of MICP-treated soils (Wang et al., 2019b).

Following the work of Al Qabany et al. (2012, 2013) and Wang et al. (2019b), both micro-scale and macro-scale experiments were conducted in this study to demonstrate how data from micro-scale microfluidic experiments that examine the effects of retention period and cementation solution concentration on the properties of calcium carbonate crystals can be used to optimise the MICP treatment of macro-scale sand soil columns for effective strength enhancement. The sizes of crystals after each of the cementation solution injections at different retention periods was quantified in the micro-scale experiments, whereas parallel micro-scale experiments were conducted to investigate the reproducibility of $CaCO_3$ crystal properties. Cementation solutions with concentrations 0.5 M or 1.0 M were also applied to study whether the dissolution of unstable and smaller $CaCO_3$ crystals also occurred at these concentrations. Subsequently, an upscaled experiment using soil columns was conducted to investigate the effects of retention period and concentration of cementation solution on crystals properties and the resulting strength of real soils.

## Materials and Methods

### Micro-scale MICP experiments using microfluidic chips

As described in Wang et al. (2019 a,b), a microfluidic chip containing porous channels is a useful tool to study the micro-scale MICP processes. **Figure 1** demonstrates the schematic of the setup for a microfluidic chip experiment, which includes a microfluidic chip, a microscope, and a flow injection system which consists of a syringe, a pump and tubing. The design and fabrication of the microfluidic chip, as well as the detailed imaging technique, is described in Wang et al. (2019a). The microfluidic chip experiment was used to observe the formation of calcium carbonate crystals over time during MICP processes involving multiple injections of cementation solution. Magnified images from previous work (Wang et al., 2019b) are shown in Figure 2 to help identify the microfluidic chip channels, bacteria and crystals.

In this micro-scale experiment, the multi-injection MICP processes involved a single injection of bacterial suspension followed by twelve injections of cementation solution performed at different retention periods and concentrations of cementation solution. In samples 1-6, the concentration of cementation solution was 0.25 M, containing 0.25 M of $CaCl_2$, 0.375 M of urea and 3 g/L nutrient broth. In samples 1-3, cementation solution was injected 2-4 times per day, with a retention period of 3-5 hours over a total period of 4 days. In samples 4-6, cementation solution was injected only once per day over 12 days. The images of samples 1 and 4 taken at the completion of the MICP processes were presented in Wang et al. (2019b), whereas more detailed results such as the images taken after each of the injections of cementation solution are shown in this study to describe the processes of the MICP in these two retention period cases in details. Samples 2, 3 and 4,5 were conducted to investigate the repeatability of the results of samples 1 and 4, respectively. Experiments with samples 7 and 8 were conducted to test whether the crystal dissolution observed using a 0.25 M solution (Wang et al., 2019b) could also occur when the concentrations of cementation solution were either 0.5 M or 1.0 M. The parameters of bacteria, bacterial injection, cementation solution and the injection of cementation solution in the three experiments are summarised in **Table 1**. The microfluidic chip experiments were conducted at room temperature.

### Macro-scale MICP experiments

*Sample preparation*

Macro-scale MICP experiments were conducted by using the setup similar to Al Qabany et al. (2012). The schematic of the experiment is shown in Figure 1 (b). Syringes with a length of 12 cm and a diameter of 35.4 mm were filled with sand and injection was achieved via gravity. After the completion of the injections, the outlet tube was bent upwards to keep the liquid inside the column for MICP reactions. The granular material being used was poorly graded sub-rounded sand with a $d_{10}$ value of 165 μm, a $d_{90}$ of 250 μm and specific gravity of 2.65 (Al Qabany et al., 2012, 2013). Each column was filled with 180 g of sand and was vibrated to achieve a final density of 1.65 g/cm³ and a porosity of about 0.37.

The parameters of bacterial suspension, bacterial injection, cementation solution and the injection of cementation solution in this macro-scale experiment are summarised in **Table 2**. Six combinations of MICP treatment conditions (concentrations and retention period) were applied in soil columns in triplicates, giving a total of eighteen columns at 22 ± 2 °C. Although the concentration of cementation solution varied (see Table 2), the total mass of cementation solution injected in terms of the available reactants was kept constant across tests by applying more injections at low concentrations and fewer injections at higher concentrations (see Table 2). The retention period was selected to maintain a total treatment duration of 6 days in treatment conditions 1-3 and 12 days in conditions 4-6.

*Unconfined compression strength (UCS) tests*

Upon completion, the MICP-treated sand samples were flushed with two pore volumes of DI water to wash away all excess soluble salts prior to drying the sand samples at 100.5°C for at least 24 hours. The top and bottom of the samples were trimmed to remove potentially disturbed or uneven zones. The UCS experiments were conducted following the ASTM D2938-86-standard test method for intact rock core specimens. The axial load was applied at a constant rate of 1.14 mm/min. The length of the sample was measured before UCS tests, and the height to diameter ratios were about 2:1 and any deviations were corrected based on Equation 3 as suggested by the ASTM D2938-86-standard test method.

$$C = \frac{C_a}{0.88 + (0.24 \, D/H)} \qquad 3$$

where $C$ is the computed compressive strength of an equivalent $H/D=2$ specimen; $C_a$ is the measured compressive strength; $D$ is the core diameter; and $H$ is its height.

*Assessment of $CaCO_3$ content and chemical efficiency*

The calcium carbonate ($CaCO_3$) content of MICP-treated soil samples was determined using the standard test method for rapid determination of carbonate content of soils (ASTM, 2004). $CaCO_3$ reacts with HCl and generates $CO_2$ (Equation 4), increasing the pressure inside a closed chamber. The actual amount of $CaCO_3$ was calculated based on a calibrated relationship (Equation 5) correlating the $CO_2$ pressure and the amount of pure analytical grade $CaCO_3$ powder (ASTM, 2004).

$$CaCO_3 + 2HCl \rightarrow CaCl_2 + CO_2 \uparrow + H_2O \qquad 4$$

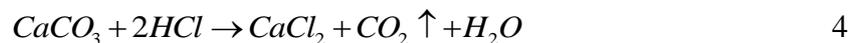

$$CaCO_3 \; content \; (g) = pressure \times 1.922 + 0.011 \qquad 5$$

The chemical efficiencies were evaluated based on the relationship between $CaCO_3$ content and the amount of chemicals injected. The concentration of urea was 1.5 times higher than the concentration of $CaCl_2$ (Martinez et al., 2013) to ensure efficient calcium transformation.

*Scanning Electron Microscopy (SEM)*

To characterise the shapes, size and distribution of precipitated $CaCO_3$ crystals inside the soil specimen, scanning electron microscope (SEM) images of MICP-treated soil samples were captured after the UCS test using a Philips XL20 scanning electron microscope (Philips Electron Optics, Eindhoven, The Netherlands). Samples were dried in an oven at 100.5°C for 24 hours. Images were taken at 300× magnification.

# Results of micro-scale experiments

## Short retention period experiment (Experiment 1)

To observe the $CaCO_3$ precipitation process in the short retention period experiment (Experiment no. 1), images of one of the middle pores in the microfluidic chip taken at the completion of the retention period of each of the cementation solution injections are shown in Figure 3a. The crystals formed after the first injection were large ones and remained present after the final injection (indicated by arrows in the first and $12^{th}$ images of Figure 3a). The small crystals shown after the $12^{th}$ injection were mainly formed after the second injection of cementation solution. In general, the crystals continued growing once formed.

The growth of the crystals in the images was assessed by measuring the diameters of three randomly selected crystals, as shown in the inset photo in Figure 3b. Crystal volumes were calculated based on an assumption that they are half-spherical (Wang et al., 2019b; Kim et al., 2020). These three crystals formed after the second injection of cementation solution. They grew after subsequent injections of cementation solution, with their growth rate being higher during the time interval between the $2^{nd}$ and the $5^{th}$ injections than between the $5^{th}$ and the $12^{th}$ injections. The decreasing growing rate of these crystals might be due to the formation of other new crystals, which consume the injected $CaCl_2$. The sizes of the three crystals were between 30 and 60 $\mu m^3$ after the second injection of cementation solution, increased by about 4-5 times after the $5^{th}$ injection of cementation solution, and became between 220 to 350 $\mu m^3$ after the $12^{th}$ injection of cementation solution. The growth rate and the final sizes of the three crystals varied. Crystal growth was affected by many factors such as the local concentration of bacterial cells and the local concentration of urea and $CaCl_2$, which are time-dependent and are affected by the existence and growth of the surrounding crystals.

Figure 4a shows the images of the middle squares with areas of 2 mm by 2 mm of microfluidic chip No. 1, 2 and 3, taken after the retention period of the final $12^{th}$ injection. Magnified images of the central pores of the squares in the images shown in Figure 4a are presented in Figure 4b. The results of the three samples are consistent, generally showing small $CaCO_3$ crystals coating the surface of the chip. The crystals formed after MICP treatment with short injection intervals (Experiment 1) were 5 - 10 µm in size, spherical in shape, with 200 - 1000 crystals present per $10^6$ µm$^3$. As stated earlier, small crystals remained inside the pores because a 3-5 hr retention period was not long enough for small crystals to dissolve.

## Long retention period experiment (Experiment 2)

Experiment 2 had a retention period of 24 hours, which was six times longer than that of Experiment 1. To observe the precipitation process of the $CaCO_3$ crystals in Experiment 2, images of the same pore in the same microfluidic chip (microfluidic chip numbered as No.4 in the present study) taken at the completion of the retention period of each of the cementation solution injections are shown in Figure 5a. Three large crystals formed in the pore after the first injection of the cementation solution and then continued growing after subsequent injections. New crystals were formed after the $4^{th}$ injection, and the sizes of the new crystals increased with injections. At the completion of the retention period, the crystals were larger but fewer in number compared with the crystals formed at the same stage in microfluidic chip No.1 (shown in Figure 3a). The crystals formed after the first injection of cementation solution were rhombohedral (Figure 5a), which is consistent with the shape of calcite. However, the morphology of crystal 1 after the sixth injection and the morphology of crystals 2 and 3 after the second injection of cementation solution is not exactly rhombohedral. Similar phenomena were observed by Mitchell and Ferris (2006) that crystal morphology appeared poorly ordered and a stepped surface topography resulted in somewhat rounded crystal edges. New crystals formed between the 6$^{th}$ and 12$^{th}$ injections of cementation solution were rhombohedral. The rhombohedral shape of calcium carbonate resembles that of calcite, which is the most stable form of $CaCO_3$. This indicated that when the retention period is longer, the $CaCO_3$ formed was more stable.

The increase in crystal size with respect to injection number is presented in Figure 5b. From the first to the 12$^{th}$ injection of cementation solution, the average crystal size increased from about $1 \times 10^4$ µm$^3$ to about $5\text{-}8 \times 10^4$ µm$^3$, which is much higher than the 220 to 350 µm$^3$ size observed after the 12$^{th}$ injection of cementation solution in the short retention period experiment. The growth rate decreased after about the 6$^{th}$ injection, which is similar to the observations made in Experiment 1. The transition to a slower rate after about the 6$^{th}$ injection could be due to bacterial activity decreasing, bacteria becoming entombed in the calcite crystals, or the other crystals which formed after the 6$^{th}$ injection of cementation solution consuming the cementation solution.

Consistent with the results shown by Wang et al. (2019b), unstable $CaCO_3$ crystals dissolved at the expense of the growth of more stable crystals after the 2$^{nd}$ and 3$^{rd}$ injections of cementation solution (Figure 6). The dissolution of the unstable crystals occurred between the 4$^{th}$ and 24$^{th}$ hour after each of the injections. A 24-hour retention period resulted in smaller spherical crystals being dissolved, and the dissolved calcium and carbonate ions from the small crystals formed bigger crystals.

Images of the middle squares with areas of 2 mm by 2 mm of microfluidic chip No. 4 to 6, taken at the completion of the last injection, are shown in Figure 7a. Magnified images of the central pores of the squares are shown in Figure 7b. The sizes of crystals are larger (10 - 80 µm) than those in microfluidic chip No. 1-3 in Experiment 1. The crystals in samples 5 and 6 are rhombohedral, consistent with the shape of calcite. The number of crystals inside the pores is small (5 - 20 per $10^6$ µm$^3$), which is about 40 times smaller than the number of crystals observed in microfluidic chip No. 1-3. Although the properties of crystals in the three samples varied, each sample showed a similar trend in crystals size, shape and distribution.

## Effect of higher concentrations of cementation solution (Experiment 3)

The concentrations of cementation solution normally used for MICP treatment are between 0.25 M and 1.0 M. Another microfluidic chip experiment (Experiment 3) was conducted to investigate whether the dissolution of smaller or relatively unstable crystals occurs at the expense of larger or more stable crystals when the concentration of cementation solution was 0.5 M or 1.0 M, which is larger than in the previous two experiments (0.25M). A long retention period of 24 hours was used for each injection. Microscope images of 250 µm × 250 µm at 1, 3, 6 and 24 hours after the completion of the second injection of cementation solution for the 0.5 M case are shown in Figure 8. Similar to what was observed in the 0.25 M case, small crystals also formed after the second injection of cementation solution and subsequently dissolved to then be replaced by larger crystals.

The dissolution of unstable crystals at the expense of growth of more stable crystals was also observed after the first injection of cementation solution when the concentration of cementation solution was 1 M (Figure 9). During the first hour, irregular-shaped $CaCO_3$ crystals and spherical $CaCO_3$ crystals were observed in the pore. The irregularly-shaped $CaCO_3$ dissolved and spherical $CaCO_3$ crystals continued to grow during the next 3 hours. At 6 hours, three more crystals appeared, while the previously observed regular-shaped $CaCO_3$ almost disappeared, and the spherical $CaCO_3$ was larger compared to its size at 3 hours. By 24 hours, the spherical $CaCO_3$ crystal dissolved while the more stable three rhombohedral crystals became larger. This result is consistent with the observations made using the other concentrations described. This process can be explained by Ostwald ripening, which is a spontaneous process driven by chemical potential differences among different-sized particles. Specifically, larger crystals grow at the expense of smaller ones which have a higher solubility than the large ones (Zhou et al., 2018).

In summary, the growth of more stable crystals at the expense of the dissolution of less stable crystals seems to occur during MICP processes, regardless of whether the concentration of cementation solution is 0.25 M, 0.5 M or 1.0 M. Therefore, in general, if the retention period between injections does not result in a decrease in bacterial activity, a longer retention period between injections results in more stable $CaCO_3$ crystals being produced.

## Results of macro-scale experiments

The micro-scale experiments showed that retention period affects the micro-scale properties of the $CaCO_3$ crystals such as size, shape, number, and stability. These micro-scale properties of $CaCO_3$ crystals significantly affect the macro-scale mechanical properties of MICP-treated sands, including their strength and stiffness. Therefore, after performing micro-scale MICP experiments, macro-scale experiments were conducted to explore the effect of retention period on the mechanical properties of MICP-treated sand. Six different MICP treatment conditions with combinations of various concentrations of cementation solution and injection intervals were used, as listed in Table 2. The total cementation solution mass was the same across the experiments. The total injection duration for short interval conditions 1, 2 and 3 was 6 days, whereas for short interval conditions 4,5 and 6 the total injection duration was 12 days. The concentration of cementation solution, injection number, and injection intervals were varied accordingly.

### Unconfined compressive strength (UCS)

Images of one of the triplicate specimens before USC tests are shown in Figure 10a. Samples processed using conditions 1, 2, 4, and 5 (0.25 M and 0.5 M) were intact when extracted from the syringe moulds, while specimens processed using conditions 3 and 6 (1.0 M) broke at weakly cemented spots. A photograph of an MICP-treated specimen in a UCS test is shown in Figure 10b, and a photograph of the sample after the UCS test is shown in Figure 10c. All samples failed with a tensile-like failure, as in previous research (van Paassen et al. (2010), Al Qabany et al. (2012) and Cheng et al. (2012)).

Figure 11a shows the UCS values plotted against $CaCO_3$ content. For comparison, the results of Al Qabany and Soga (2013) are also shown. UCS varied from 0 to 5.5 MPa at the same $CaCO_3$ content. These large strength variations at the same cementation level is consistent with the results reported by Wang et al. (2017) because the $CaCO_3$ crystals have different characteristics depending on the MICP treatment conditions. In general, higher UCS values were obtained when the retention period was longer. For the same level of $CaCO_3$ content produced at 0.25 M, 0.5 M and 1.0 M concentrations of cementation solution, the UCS values increased by about 4.3, 5.8 and 3.2 times, respectively, when the treatment duration increased from 3 days (Al Qabany and Soga 2013) to 6 days (this study) (Figure 11a); average UCS values increased by a further 28%, 27% and 13%, respectively, when the total treatment duration increased from 6 days to 12 days (Figure 11b). The increase in treatment duration from 3 days to 6 days resulted in a large number of smaller crystals dissolving and reprecipitating into larger crystals, thereby bonding the soil particles more efficiently and thus significantly increasing soil strength.

The UCS of samples treated with 1.0 M cementation solution are in general about half of the values of samples treated with cementation solution of 0.25 M or 0.5 M when the total duration is 6 or 12 days. This finding is also consistent with the results presented in Al Qabany and Soga (2013) reporting low UCS values of samples treated with cementation solution of 1.0 M. The low UCS values at 1 M is mainly because of the inhomogeneity of samples (Al Qabany and Soga (2013)). Since the 1.0 M samples broke during the extraction process (Figure 10a), the UCS values of samples treated with 1.0 M cementation solution shown in Figure 11a are calculated by testing the larger part of the two broken parts of each of the samples. Thus, the strengths of the samples obtained in this experiment are higher than the strength obtained in the study of Al Qabany and Soga (2013).

## Chemical efficiency

The chemical efficiencies of the samples are shown in Figure 12. When the concentration of cementation solution was either 0.25 M or 0.5 M, the chemical efficiency was relatively high (higher than 75 %). These results are consistent with the study by Konstantinou et al. (2020), where larger specimens of 70 mm diameter were generated with a similar setup and the same urea to calcium chloride ratio as in this study (1.5:1). By contrast, when the concentration of cementation solution was 1.0 M, the mean chemical efficiency was lower (71 % and 64% for 6 day treatment and 12 day treatment, respectively), since the long retention period (4 days) likely caused a decrease in bacterial activity due to the higher molarity entombing some of the bacteria over time. However, the bacterial activity inside the soil was difficult to measure.

The variation in chemical efficiency between samples treated with the same MICP procedure, shown by the error bars in Figure 12, was large when the cementation solution concentration was 1.0 M. The variations in efficiencies were largest for 1.0 M - 6 d and 1.0 M - 12 d at about 42% and 16%, respectively. The large variations in efficiencies indicated the inhomogeneity of soil samples, which is consistent with the results obtained by Al Qabany et al. (2012).

Since the size and shape of the crystals varied, it was difficult to correctly quantify the chemical efficiency in the microfluidic chip experiments based on the number and sizes of crystals when irregular shapes or clusters were formed. This resulted in calculated chemical efficiencies in the microfluidic chip experiments of about 35% and 40% in the short retention period and long retention period cases, respectively, which were lower compared to the ones in the soil column experiments.

## Micro-scale properties of $CaCO_3$ crystals observed by SEM

An increase in retention period between injections can increase the performance efficiency since the microfluidic experiments show that the micro-scale properties of $CaCO_3$ are affected by retention period. A smaller number of larger and more stable $CaCO_3$ is formed when the retention period increases. To observe the $CaCO_3$ crystals after the MICP treatment of the macro-scale specimens, scanning electron microscope (SEM) images were taken (see Figure 13). When the concentration of cementation solution is the same, the crystals in the samples treated with a longer retention periods are larger than those with shorter retention periods. For example, in the case of 0.25 M concentration, the average size of the crystals increased from 30 µm at a 12-hour retention period (Figure 13a) to 50 µm at a 24-hour retention period (Figure 13b). In contrast, the average size of the crystals at 3-hour retention was about 5 µm (Al Qabany et al., 2012). These results are in agreement with the results obtained in the micro-scale experiments (Figures 3 and 5); when the concentration of cementation solution was 0.25 M, the crystal size after the 12$^{th}$ injection of cementation solution increased from about 10 µm at the retention period of 4 hours to 50 µm at the retention period of 24 hours.

This work shows that the samples with a higher strength tend to have larger $CaCO_3$ crystals, which is consistent with the findings of Cheng et al. (2012). A possible explanation for the effect of crystals size on increasing the strength of MICP-treated samples might be that crystals large enough to fill the gaps between soil particles can prevent particle rotations during shearing, providing more resistance to dilation (Zhao et al., 2018).

## Conclusions

This study demonstrates that MICP treatment parameters for strength enhancement (macro-scale) can be designed based on micro-scale microfluidic experiments. The effects of injection interval and concentration of cementation solution on the properties of calcium carbonate crystals were examined at both scales. The findings of this study are summarised as follows.

Both the micro-scale microfluidic chip experiments and macro-scale column tests indicated that when the retention period was shorter, i.e. 3-5 hours compared to 24 hours for a 0.25 M cementation solution, the resulting crystals were larger in number and smaller in size. In addition, the micro-scale experiments showed that large crystals grew at the expense of the dissolution of smaller crystals, regardless of whether the concentration of cementation solution was 0.25 M, 0.5 M or 1.0 M. This process can be explained by Ostwald ripening, which is a spontaneous process driven by chemical potential differences between different-sized particles, where larger crystals grow at the expense of smaller ones which have a higher solubility than the large ones.

The difference in crystal sizes and numbers substantially affected the strength of MICP- treated specimens. The UCS values of samples treated with a treatment duration of 6 days using 0.25 M, 0.5 M and 1.0 M cementation solution, were 4.3, 5.8 and 3.2 fold higher of those treated with a duration of 3 days respectively. The substantial strength increase of soils treated over 6 days compared to over 3 days was largely because between 3 days and 6 days the large number of smaller crystals dissolved and reprecipitated into larger crystals which bonded the soil particles more efficiently. UCS values increased by a further 28%, 27% and 13%, respectively, when the total treatment duration increased from 6 days to 12 days. The less pronounced increase in the strength increase of soils treated over 12 days compared to over 6 days was largely because that by 6 days the crystals were already relatively large enough to bond the soil particles efficiently and further growth crystal growth increased soil strength but to a lesser extent than the previous case.

In contrast to the soil column tests, microfluidic chip experiments show changes in crystal sizes and numbers with time and provide direct information about the MICP process. This study establishes the link between MICP micro-scale microfluidic chip experiments and macro-scale column experiments, demonstrating that microfluidic chip experiments are a powerful tool for optimising MICP-treatment to produce calcium carbonate crystals with desired properties for field applications.

**Acknowledgements**
Y.W. would like to acknowledge Cambridge Commonwealth, European and International Trust, and China Scholarship Council, which collectively funded this project. J.T.D. acknowledges the support of the Engineering Research Center Program of the National Science Foundation under NSF Cooperative Agreement No. EEC-1449501. Any opinions, findings and conclusions or recommendations expressed in this manuscript are those of the authors and do not necessarily reflect the views of the National Science Foundation.

Table 1 Parameters of microfluidic chip experiments

| Experiment No. | | 1 | 2 | 3 | |
|---|---|---|---|---|---|
| Microfluidic chip No. | | 1-3 | 4-6 | 7 | 8 |
| Bacterial suspension and injection | $OD_{600}$ | 1.0 | 1.0 | 1.0 | 1.0 |
| | Flow rate (PV/h) | 56 | 56 | 56 | 56 |
| | Injection PV | 1.25 | 1.25 | 1.25 | 1.25 |
| | Number of injections | 12 | 12 | 1 | 1 |
| | Settling duration (h) | 18-24 | | 24 | 24 |
| Cementation solution and injection | Content | 0.25 M* | 0.25 M* | 0.5 M** | 1 M*** |
| | Flow rate (PV/h) | 5.6 | 5.6 | 5.6 | 5.6 |
| | Injection PV | 1.25 | 1.25 | 1.25 | 1.25 |
| | Number of injections | 12 | 12 | 2 | 2 |
| | Injection interval (h) | 3-5**** | 24 | 24 | 24 |
| | Number of injections per day | 2-4 | 1 | 1 | 1 |
| | Total treatment duration (days) | 4 | 12 | 2 | 2 |

Note: PV-pore volume of the microfluidic chip; *0.25 M $CaCl_2$, 0.375 M urea, and 3 g/L nutrient broth; **0.5 M $CaCl_2$, 0.75 M urea, and 3 g/L nutrient broth; ***1.0 M $CaCl_2$, 1.5 M urea, and 3 g/L nutrient broth; ****It should be noted that because the injections were not conducted during out of working hours, the retention period between the last injection of one day and the first injection of the subsequent day was longer than 3-5 hours.

Table 2 Parameters of macro-scale experiments

| | | Short interval | | | Long interval | | |
|---|---|---|---|---|---|---|---|
| Treatment condition No. | | 1 | 2 | 3 | 4 | 5 | 6 |
| Soil column No. | | 1~3 | 4~6 | 7~9 | 13~15 | 10~12 | 16~18 |
| Bacterial suspension and injection | $OD_{600}$ | 1.0 | 1.0 | 1.0 | 1.0 | 1.0 | 1.0 |
| | Number of injections | 1 | 1 | 1 | 1 | 1 | 1 |
| | Injection PV number | 1.2 | 1.2 | 1.2 | 1.2 | 1.2 | 1.2 |
| | Settling duration (h) | 24 | 24 | 24 | 24 | 24 | 24 |
| Cementation solution and injection | Concentration (M) | 0.25 | 0.5 | 1.0 | 0.25 | 0.5 | 1.0 |
| | Number of injections | 12 | 6 | 3 | 12 | 6 | 3 |
| | Injection interval (h) | 12 | 24 | 48 | 24 | 48 | 96 |
| | Number of injections per day | 2 | 1 | 0.5* | 1 | 0.5 | 0.25 |
| | Total treatment duration (days) | 6 | 6 | 6 | 12 | 12 | 12 |

Note: Concentration of cementation solution were indicated by the concentration of $CaCl_2$ in the cementation solution; the concentration of urea is 1.5 times of $CaC_2$; 3 g/L nutrient broth was contained in the cementation solution; * 0.5 injection per day means 1 injection in two days.

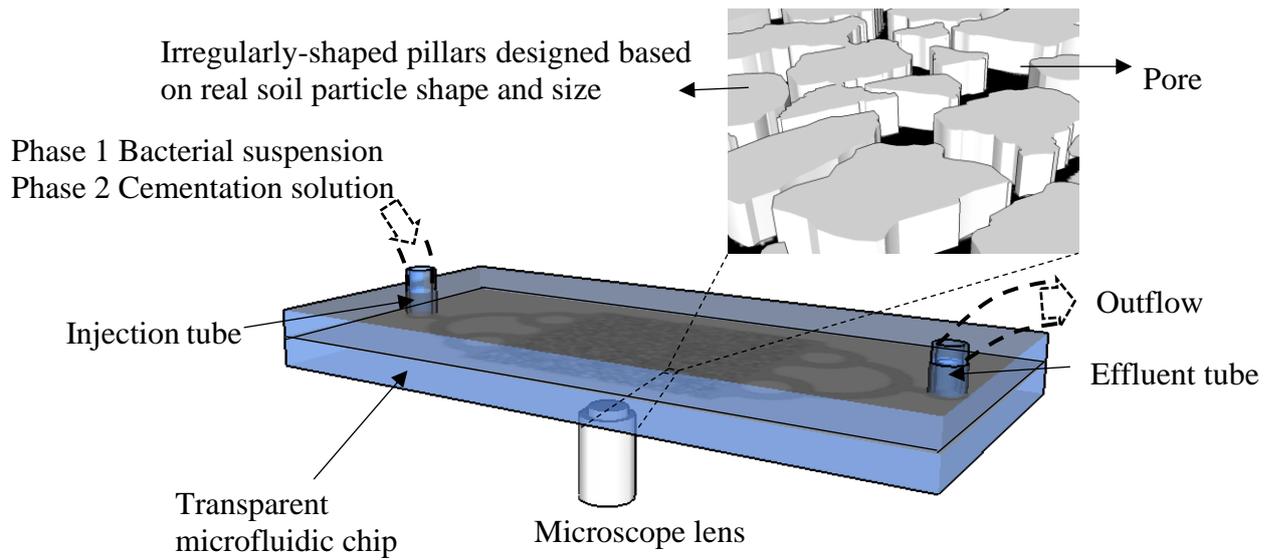

(a)

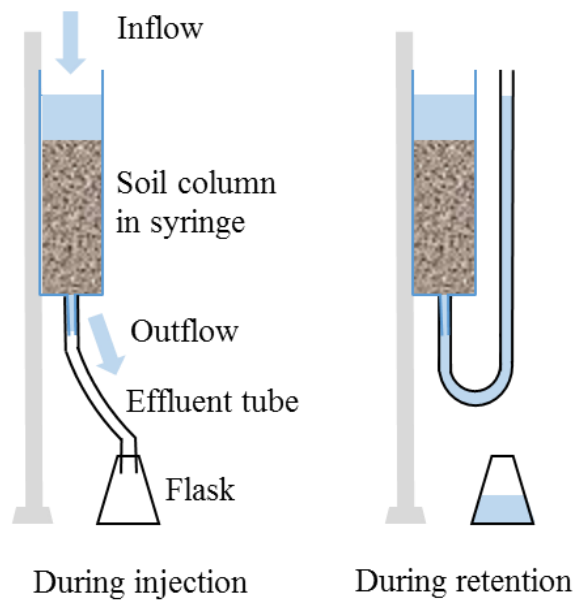

(b)

**Figure 1** Schematic of the micro and macro scale experiments. (a) micro-scale microfluidic chip experiments (Wang et al., 2019a); (b) macro-scale soil column experiments (redraw based on Al Qabany et al., 2012)

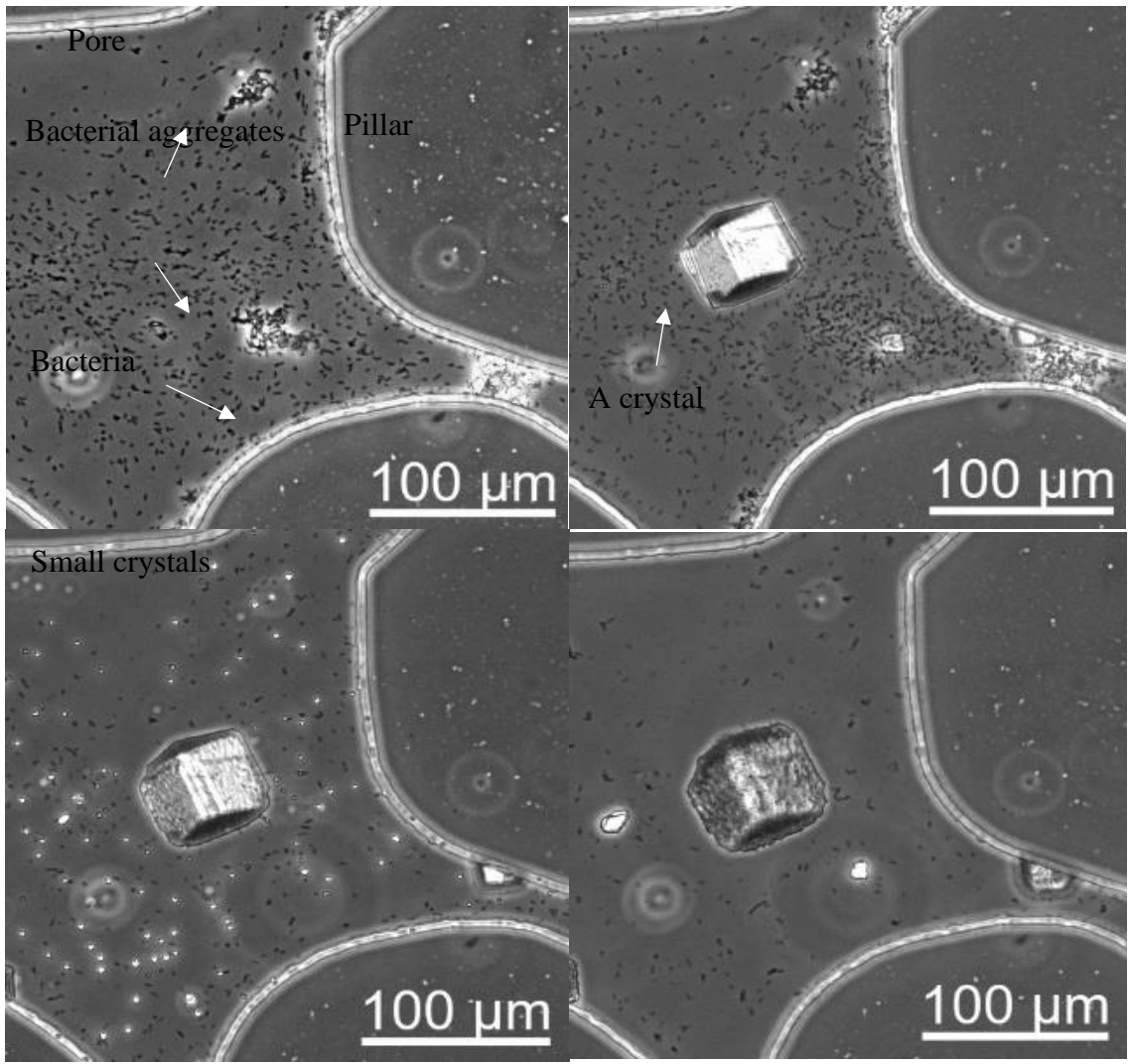

**Figure 2** Microscope images of one pore in the microfluidic chip at varies times after the injection of cementation solution as representative examples showing the properties of bacterial and calcium carbonate crystals (replotted from Wang et al., 2019b)

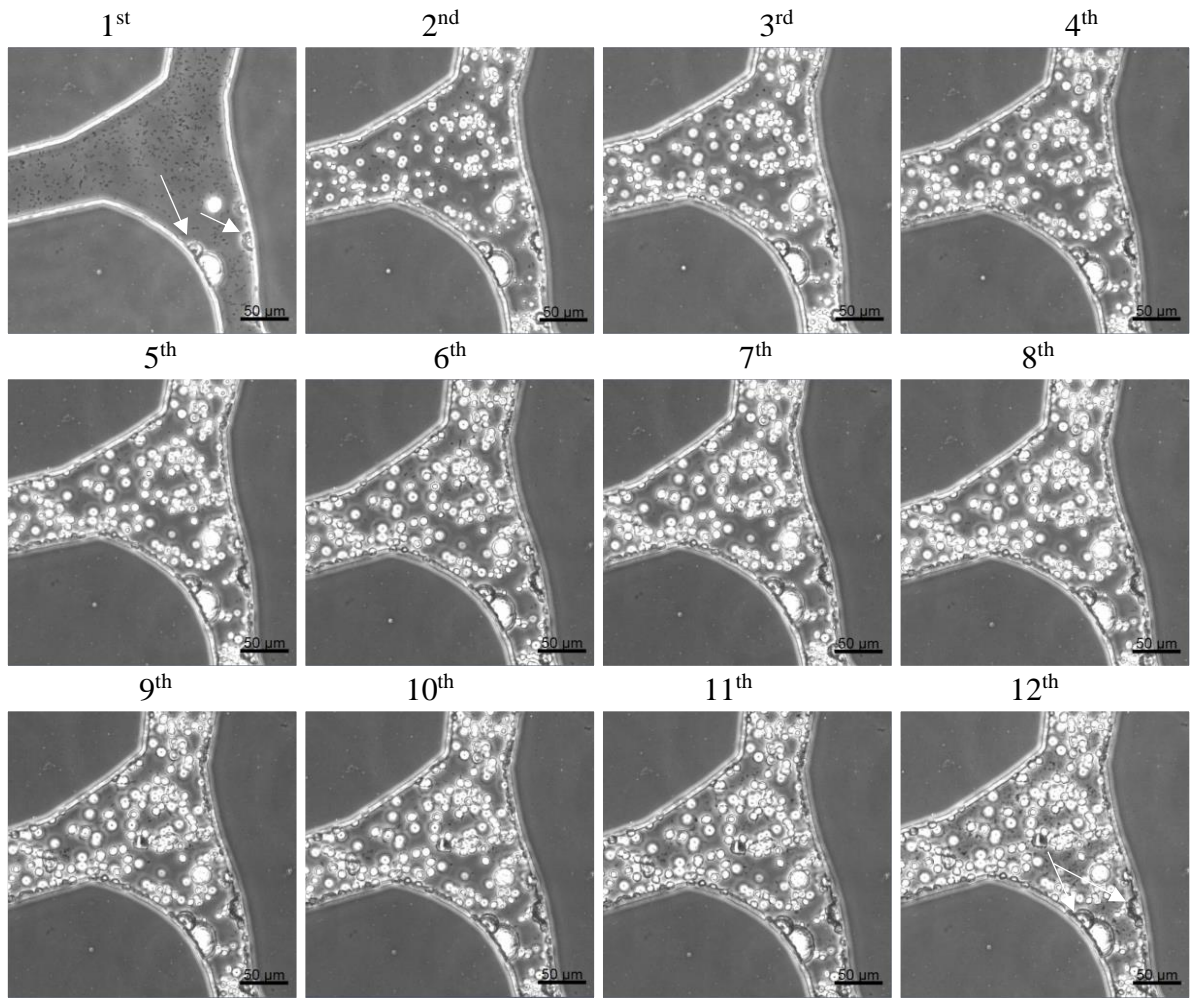

(a)

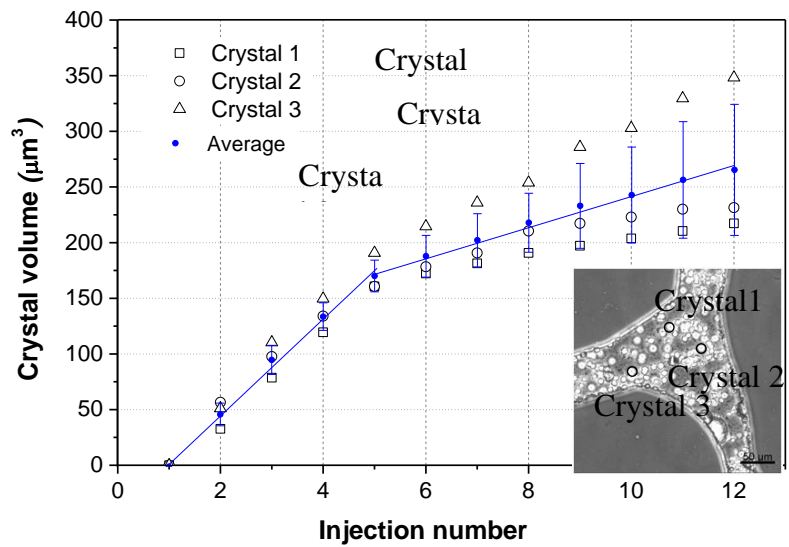

(b)

**Figure 3** (a) Microscope images of the centre pore of microfluidic chip No.1 (3-5 h injection interval) at the completion of the retention period of all the 12 injections of cementation solution; the last injection microscope image in (a) was presented in Wang et al. (2019b); (b) Increase in volume of three crystals with injection; the average volume of the three crystals was also plotted with time, and data presented as mean ± standard error;

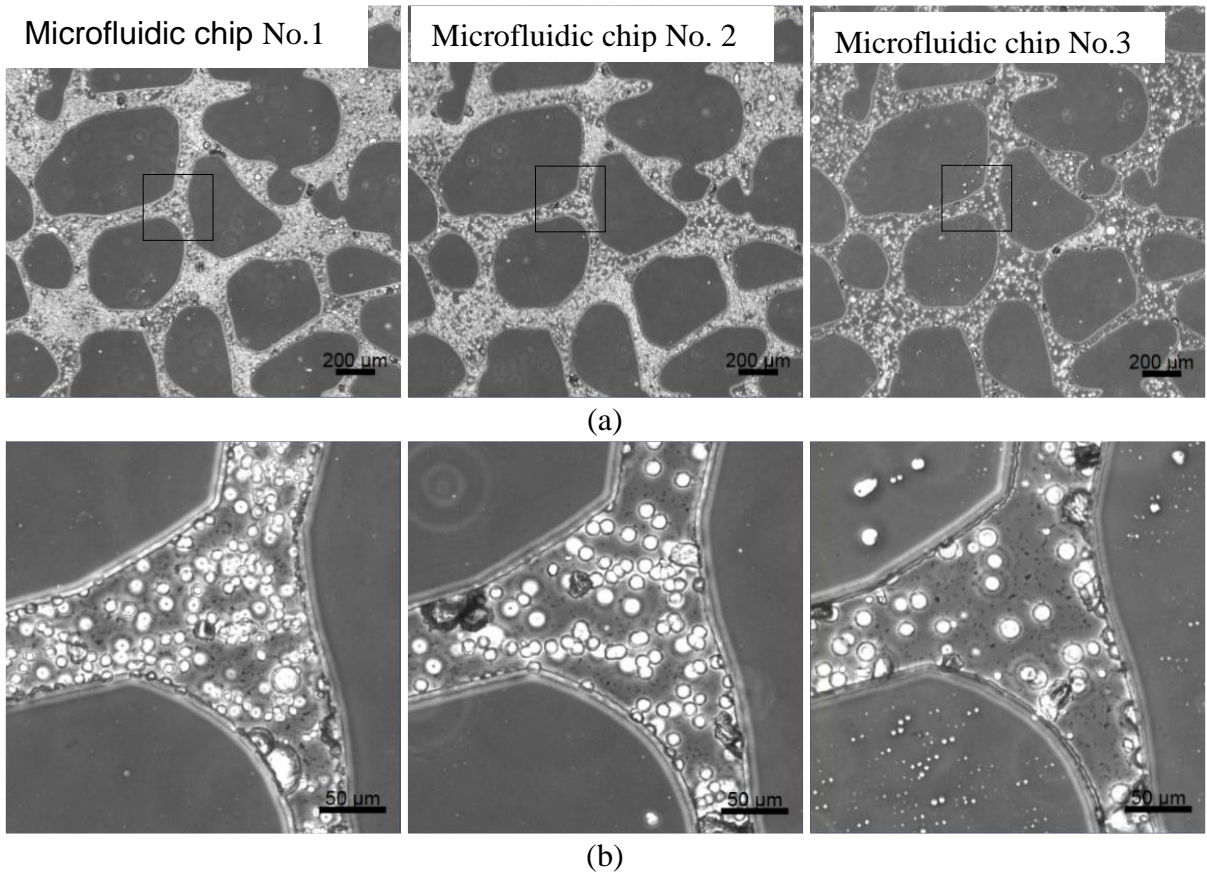

**Figure 4** Microscope images of microfluidic chip No. 1-3 (3-5 h injection interval) at the completion of the retention period of the final injection of cementation solution. (a) images of the centre 2 mm by 2 mm squares; (b) magnified images of pores marked by black squares in (a); the two images of microfluidic chip No. 1 were presented in Wang et al. (2019b)

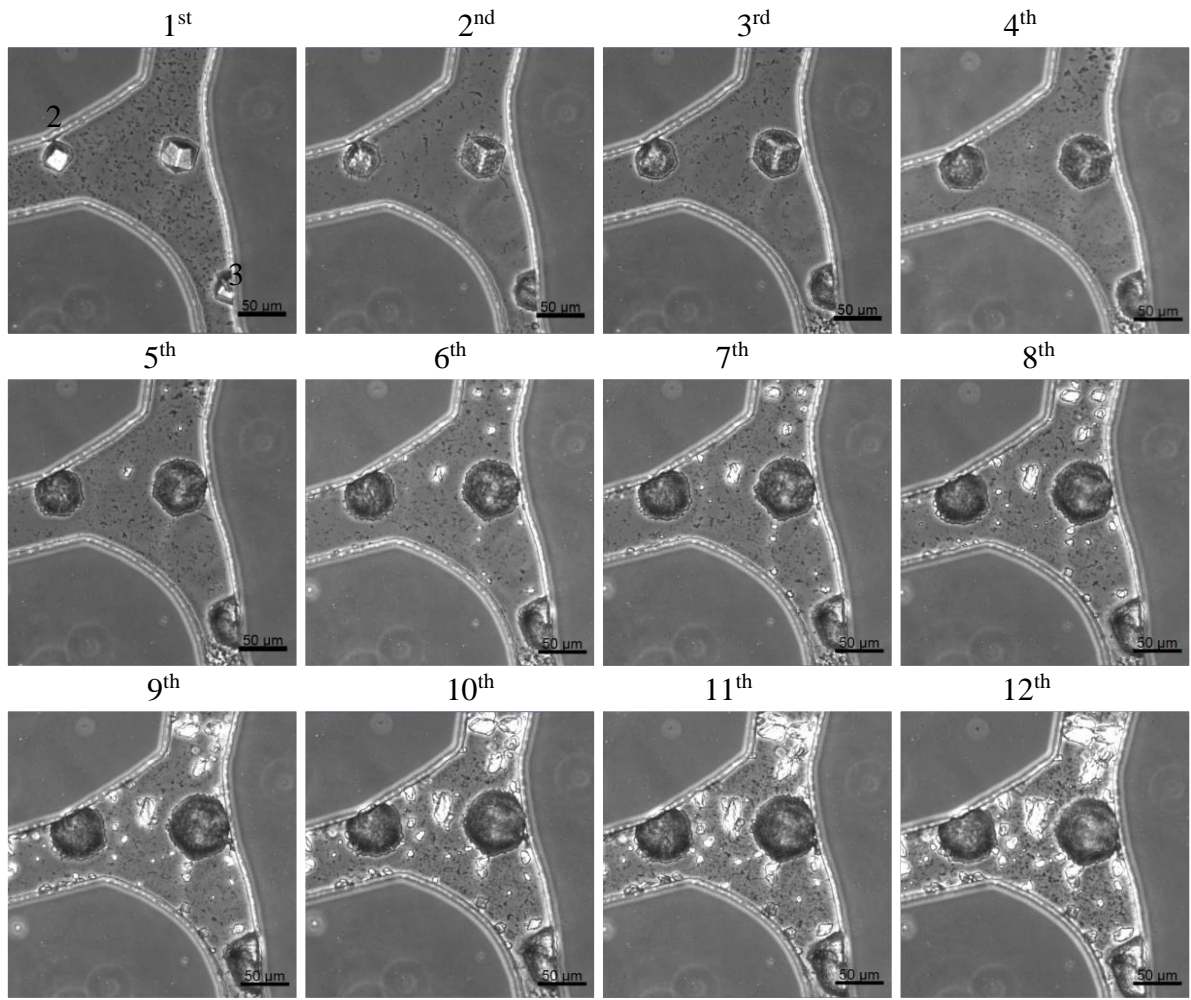

(a)

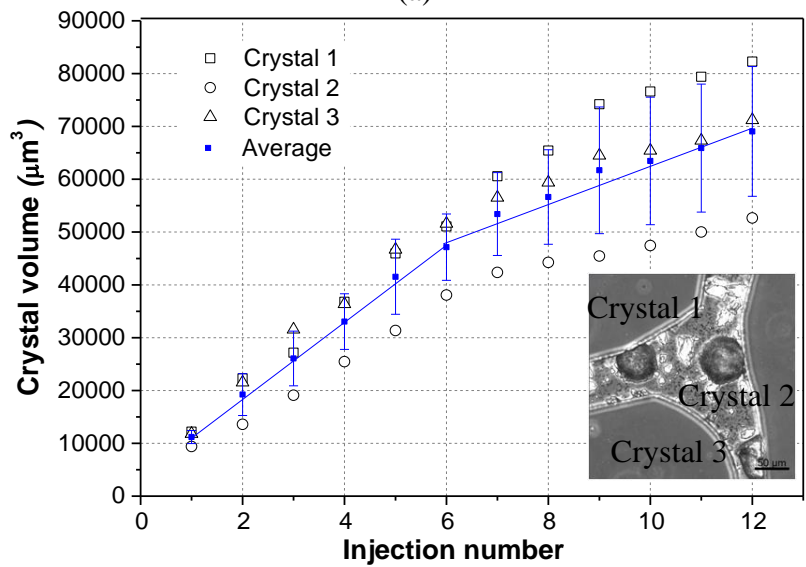

(b)

**Figure 5** (a) Microscope images of the centre pore of sample 1 (24 h injection interval) at the completion of the retention period of each injection of cementation solution; the last injection microscope image in (a) was presented in Wang et al. (2019b); (b) Crystal sizes after each injection of cementation solution; the average volume of the three crystals was also plotted with time, and data presented as mean ± standard error

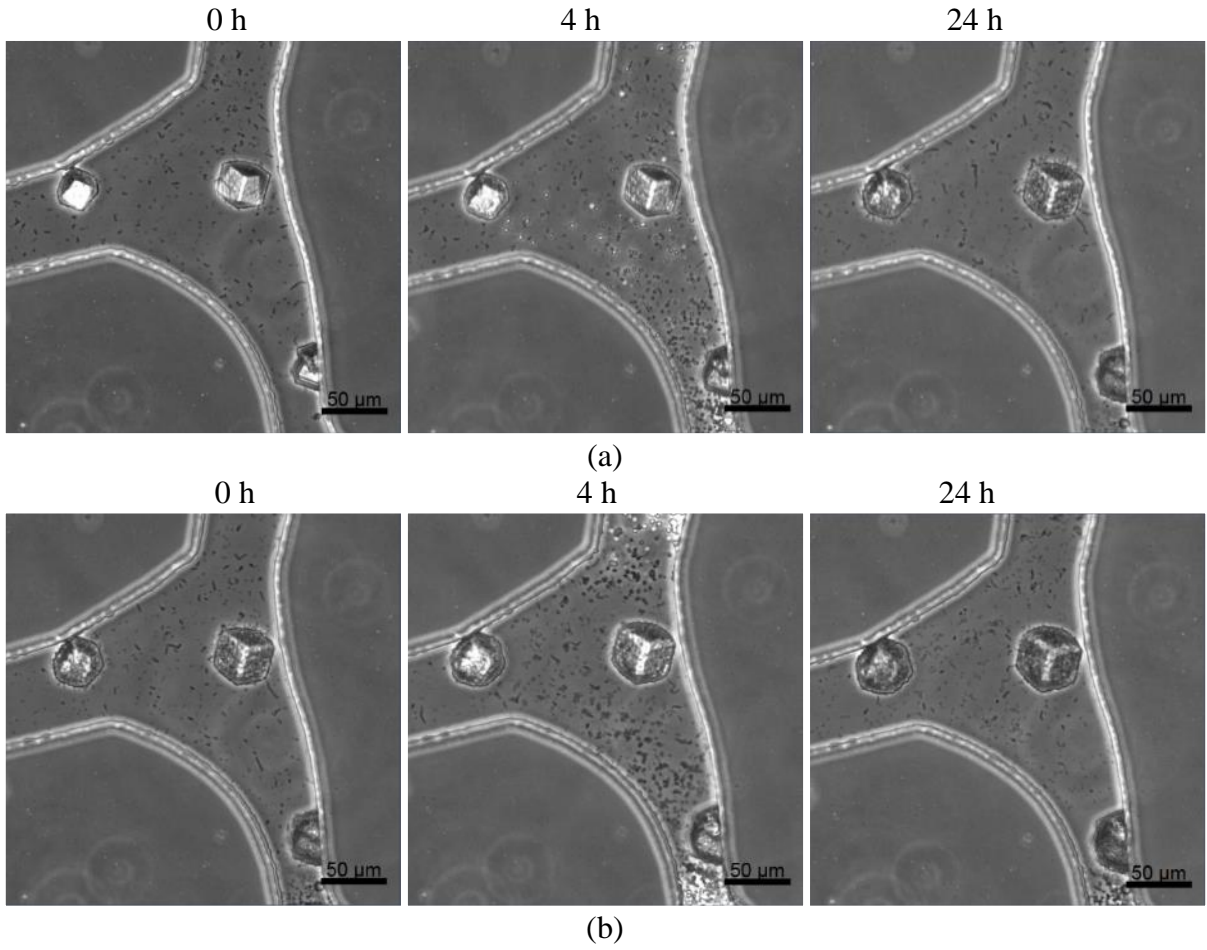

**Figure 6** Microscope images of microfluidic chip No. 4 (24 h injection interval) at 0, 4, and 24 h after the second (a) and third (b) injection of cementation solution

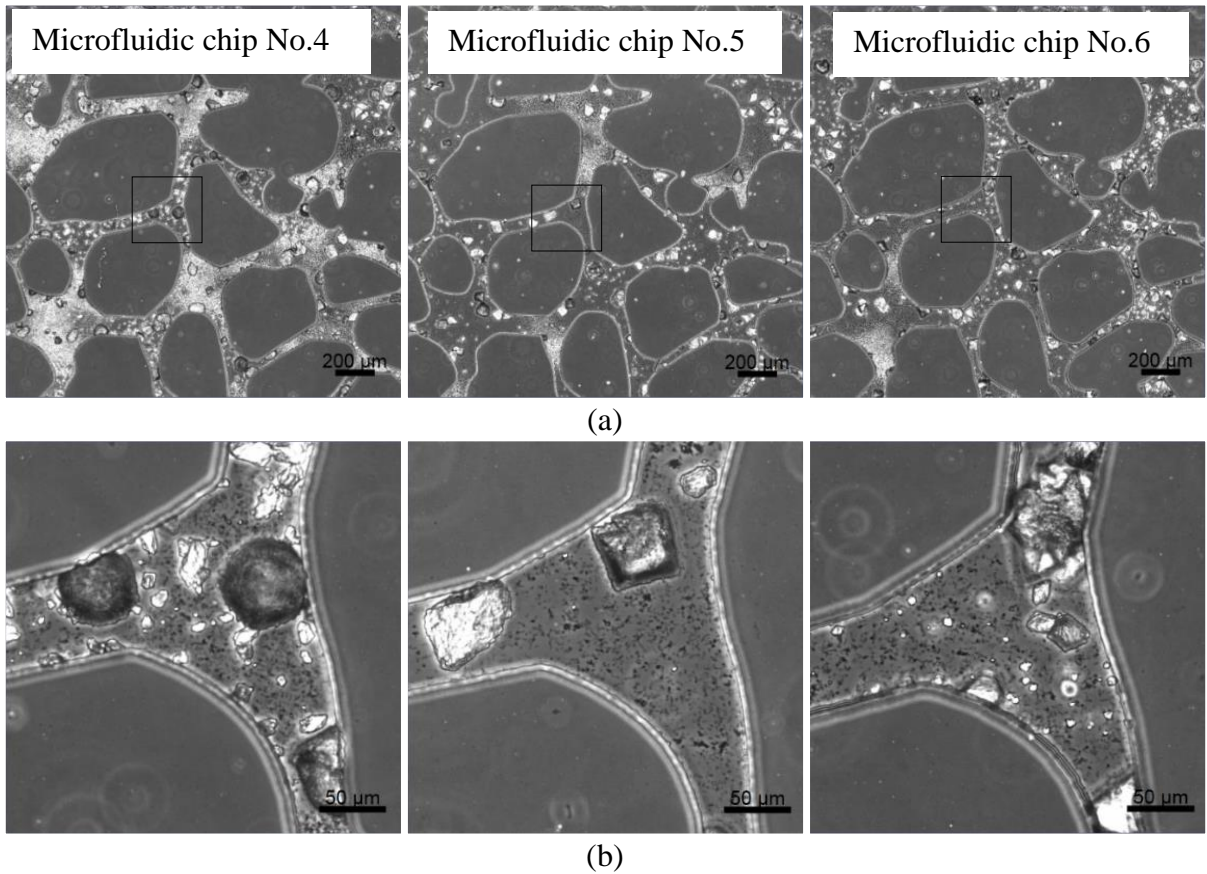

**Figure 7** Microscope images of microfluidic chip No. 4-6 (24 h injection interval) at the completion of the retention period of the final injection of cementation solution. (a) images of the centre 2 mm by 2 mm squares; (b) magnified images of the pores marked by black squares in (a); the two images of microfluidic chip No. 4 were presented in Wang et al. (2019b)

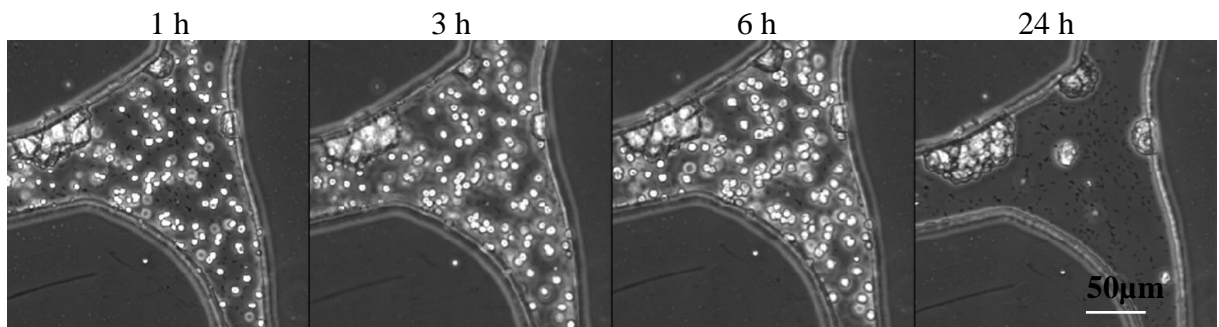

**Figure 8** Microscope images of 250 µm by 250 µm square at the centre of microfluidic chip No. 7 at 1, 3, 6 and 24 hours after the completion of the second injection of cementation solution

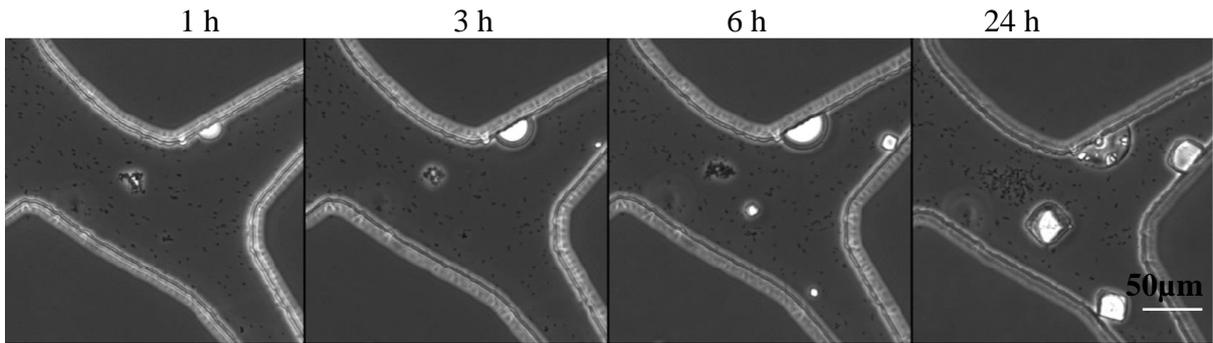

**Figure 9** Microscope images of 250 µm by 250 µm square at the centre of microfluidic chip No. 8 taken at 1, 3, 6 and 24 hours after the completion of the first injection of cementation solution

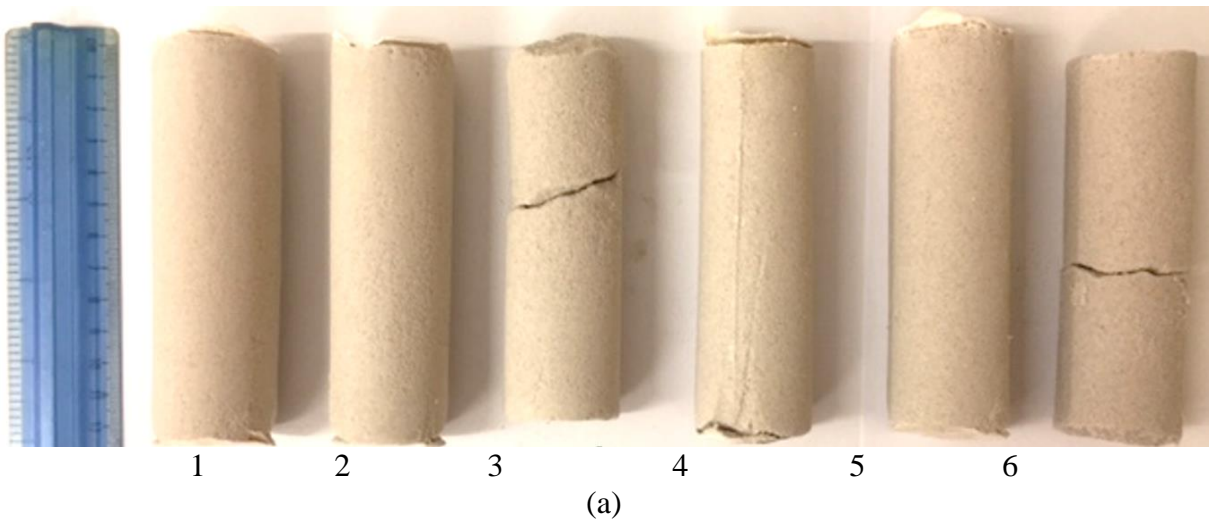

(a)

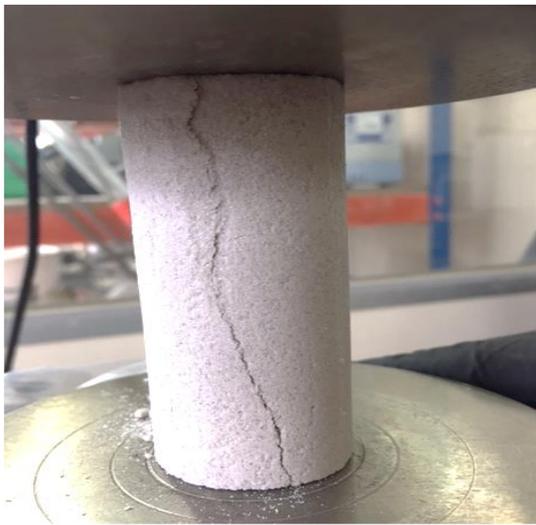

(b)

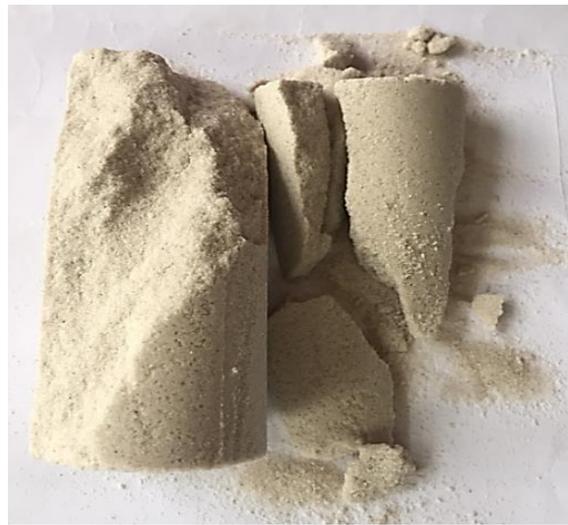

(c)

**Figure 10** (a) Exemplary photos of samples in the six macro-scale MICP experiments; (b) a photo of an MICP-treated specimen in a UCS test; (c) a photo of the sample after being broken during the UCS test

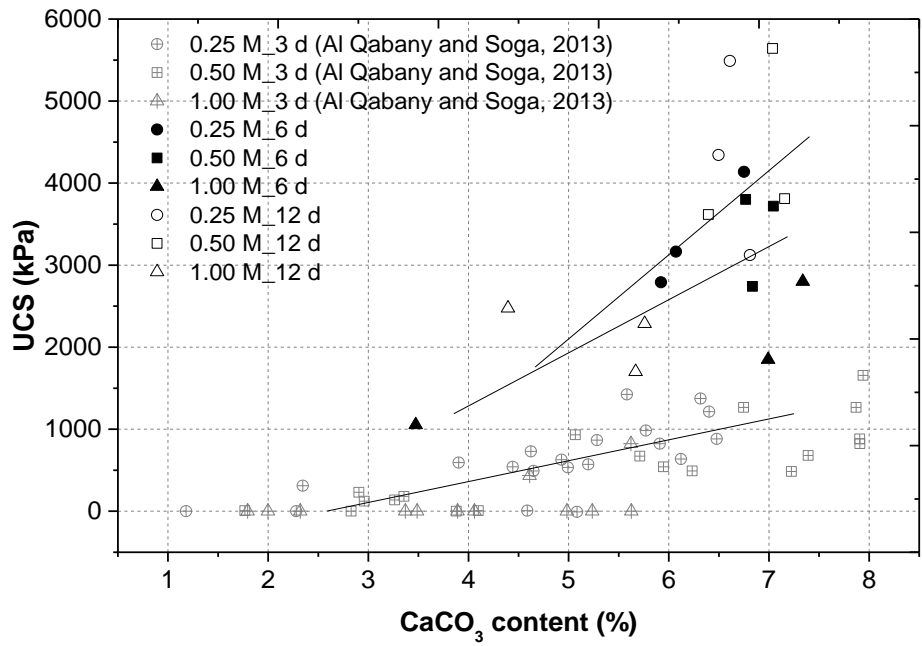

(a)

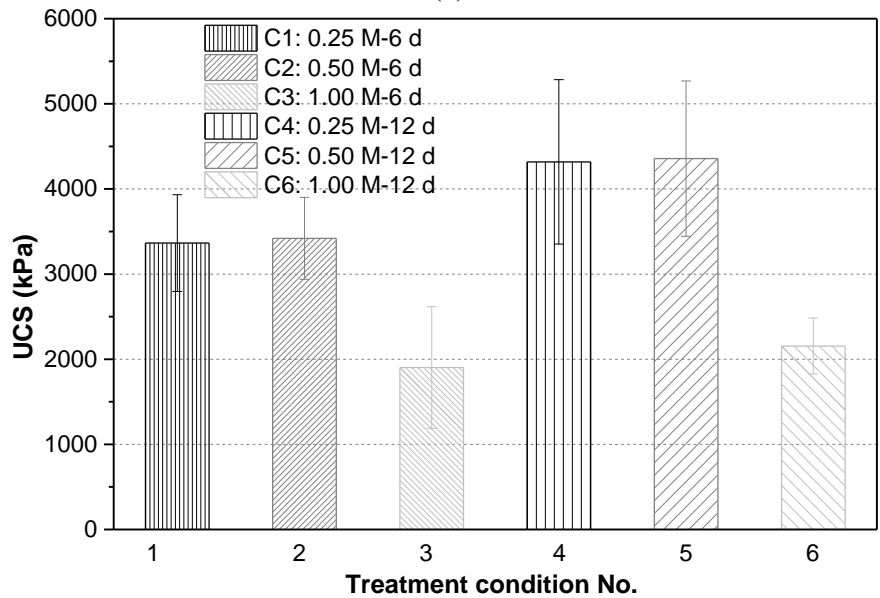

(b)

**Figure 11** (a) CaCO$_3$ content vs UCS (comparison with the results of Al Qabany and Soga, 2013); (b) Unconfined compressive strength (UCS) values of MICP-treated sand samples in this study. Data presented as mean ± standard error, n=3 (n is the number of times each treatment condition and the relative measurement was repeated)

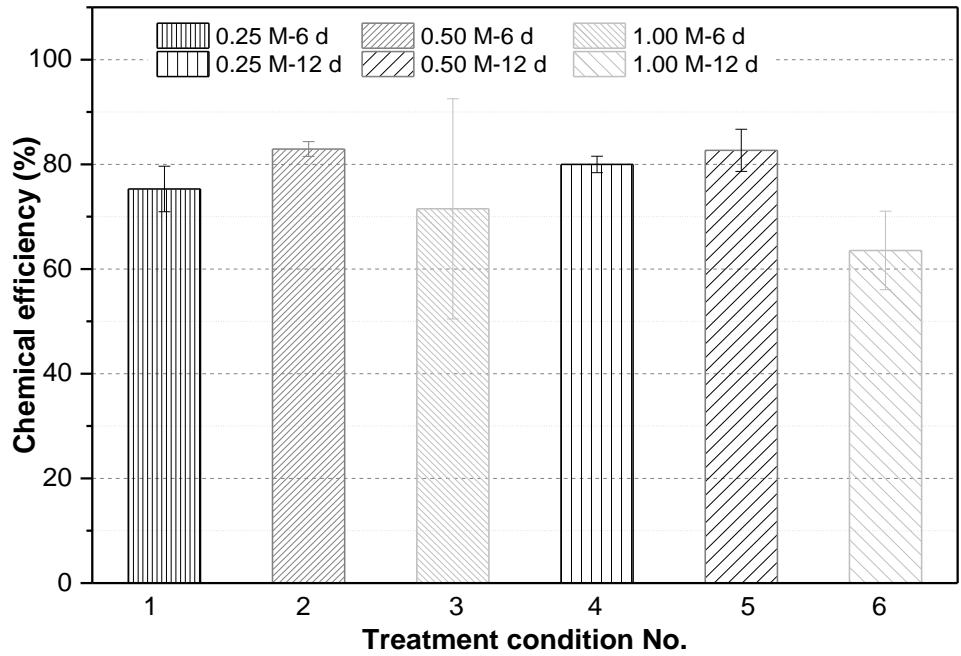

**Figure 12** Chemical efficiencies of the MICP-treated sand samples. Data presented as mean ± standard error, n=3 (n is the number of times each treatment condition and the relative measurement was repeated)

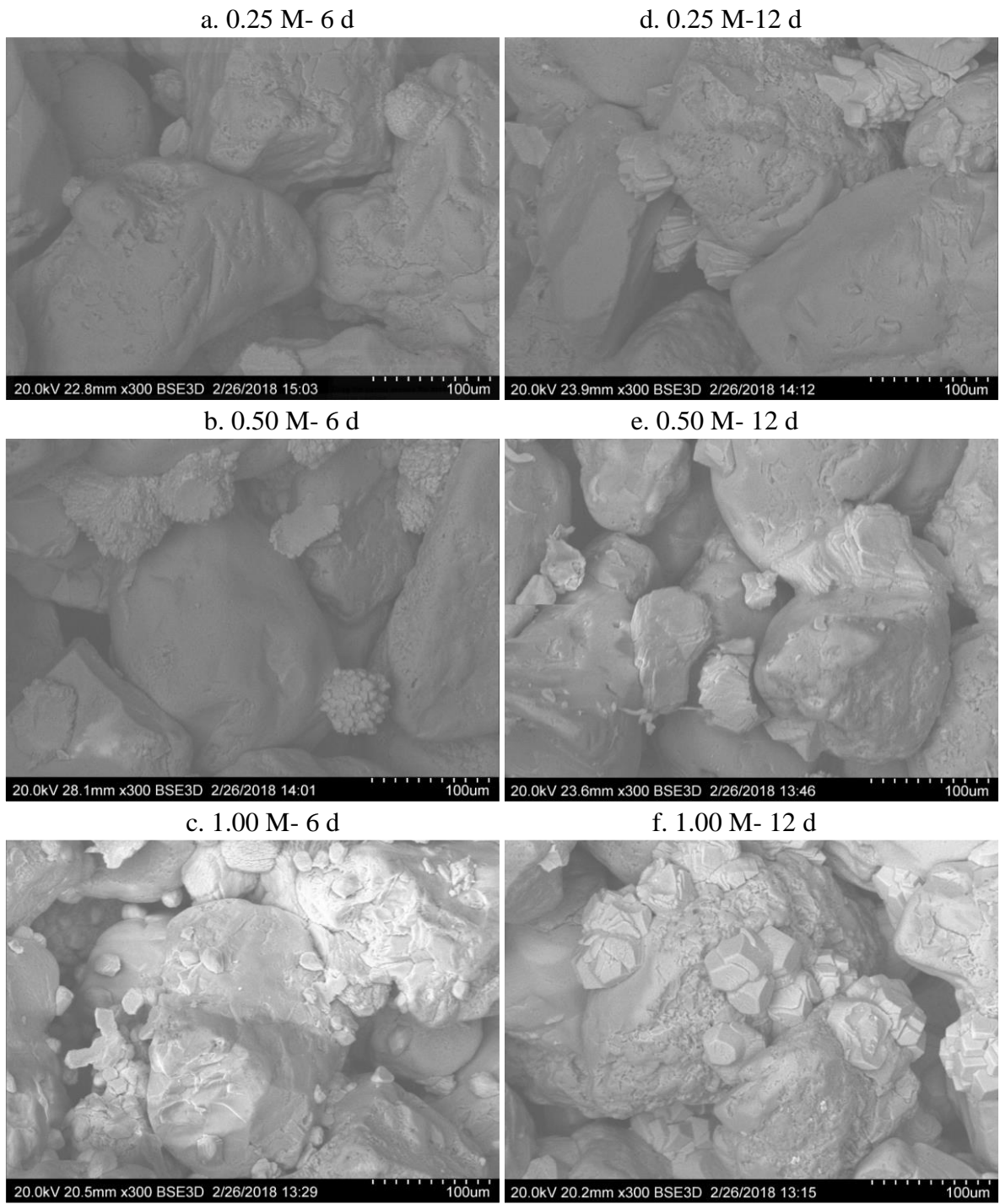

**Figure 13** SEM images of CaCO$_3$ crystals inside MICP-treated sand samples after MICP treatments. a. 0.25 M-6 day treatment, CaCO$_3$ content is 6.1 %; b. 0.50 M- 6 day treatment, CaCO$_3$ content is 7.0 %; c. 1.00 M-6 day treatment, CaCO$_3$ content is 7.0 %; d. 0.25 M-12 day treatment, CaCO$_3$ content is 6.6 %; e. 0.50 M-12 day treatment, CaCO$_3$ content is 7.0 %; f. 1.00 M-12 day treatment, CaCO$_3$ content is 5.8 %